\documentstyle[12pt]{article}
\renewcommand{\theequation}{\arabic{section}.\arabic{equation}}
\renewcommand{\thesection}{\Roman{section}.}

\begin{document}
\title{The $1/N$ Expansion and Spin Correlations in
Constrained Wavefunctions}
\author{Maxim Raykin\thanks{Email: raykin@buphy.bu.edu}\\
{\it \small Physics Department, Boston University, Boston, MA 02215,
USA}\\
and\\
Assa Auerbach\thanks{Email: assa@phassa.technion.ac.il}\\
{\it \small Physics Department, Technion, Haifa 32000, Israel and}\\
{\it \small Physics Department, Boston University, Boston,
MA 02215, USA }}
\date{}
\maketitle
\begin{abstract}
We develop a large-N expansion for
Gut\-zwil\-ler projected  spin \linebreak states.
We consider valence bonds singlets, constructed by
Schwinger bosons or fermions, which are variational
ground states for quantum antiferromagnets.
This expansion is simpler than the familiar expansions of
the quantum Heisenberg model, and thus more instructive.
The diagrammatic rules of this expansion allow us to
prove certain identities to all orders in 1/N.
We derive the on-site spin fluctuations sum rule for arbitrary N.
We calculate the correlations of
the one dimensional Valence Bonds Solid states and the
Gutzwiller Projected Fermi Gas upto order 1/N.
For the bosons case, we are surprised to find
that the mean field, the order 1/N
and the exact correlations are simply proportional.
For the fermions case, the 1/N
correction enhances the zone edge singularity.
The comparison of our leading order terms to known results for N=2,
enhances our understanding of \mbox{large-N} approximations
in general.
\end{abstract}

{\noindent\hspace{.42in}\footnotesize
}

\section{Introduction}

The use of large-N approximations to treat strongly interacting
quantum systems has been very extensive in the last decade.
The approach has
originated in particle physics, but has found many applications
in condensed matter systems. Some
notable examples are the SU(N) quantum Heisenberg model
\cite{aa,RS} the (closely related) non linear sigma
or CP$^{N-1}$ models \cite{poly,PC},
the  Anderson and Kondo (Coqblin-Schrieffer) models
\cite{ki,kl}, and the two band Hubbard model for cuprate
superconductors \cite{levin}.

Generally speaking,
the parameter N labels an internal SU(N)  symmetry at each lattice
site, (i.e. the number of ``flavors'' a particle can have). In
most cases, the large-N approximation has been applied to treat
spin hamiltonians, where the symmetry is SU(2), and N is therefore
not a truly large parameter. Here lies its primary weakness, since
in most cases the ${\rm N} > 2$ models are not physically realizable.
Nevertheless, the 1/N expansion provides an easy method
for obtaining simple mean field theories. These have been found to
be either surprisingly successful, or completely wrong depending on
the system. For example: the Schwinger boson mean field theory
works well for the quantum Heisenberg model,
except for the half-odd integer antiferromagnet in one dimension
\cite{aa,RS}. The latter is better
described by a fermion \mbox{large-N} approximation.
It is important to investigate the conditions which allow certain
large-N generalizations to yield a ``better'' mean field theory for
a particular N=2 system.

In contrast to spin wave expansions about a broken symmetry state,
the large-N approach
can describe both ordered and disordered  phases.
At ${\rm N}=\infty$, the generating functional is dominated by its
saddle point, which is a non interacting mean
field theory with few variational parameters.  The variational
equations and the leading order
correlations are in many cases analytically tractable.

The  corrections to the mean field theory are given by Feynman
diagrams, where the ``interactions'' are mediated by RPA matrix
propagators. It is hard in most cases to compute these diagrams
even to first order in 1/N, which is why they have been determined
only in a few select cases \cite{ki,kl}.

In this paper we shall start by deriving a new and
simplified version of the large-N expansion
suitable for evaluating spin correlations in constrained variational
wavefunctions. These states have been used as trial ground states
for various antiferromagnetic Heisenberg models. The exact
calculation of their correlations is not feasible in most cases.
In one dimension, two cases which have been solved
analytically are the Valence
Bonds Solids (VBS), and the Gutzwiller Projected Fermi Gas (GPFG).
We shall make use of these exact soultions in this paper.

It is our primary purpose to study the properties of the 1/N
expansion by using the constrained states
as ``toy'' problems. Their 1/N expansion
differs from that of e.g. Ref. \cite{aa}
in two respects: (i) Here, the generating functional has no
time dependence and Matsubara sums, and (ii) there is only one
fluctuating field per site, the constraint field $\lambda$, and no
Hubbard-Stratonovich fields.
Thus, we study the ``pure'' effects of the constraints,
without the interactions effects of the quartic Hamiltonian.
These features simplify the evaluation of 1/N corrections
considerably.

This paper is organized as follows. Section II introduces
the valence bonds states.
Section III introduces the Gutzwiller projected Fermi gas states.
Section IV defines the generating functional of the spin
correlation functions for both bosons and fermions states in a
unified notation. Section V derives the 1/N
expansion of the correlation functions and describes the
diagrammatic rules. Section VI applies the diagrammatic rules to
prove three identities to all orders in 1/N: the absence of charge
fluctuations, the sum rule for on-site spin fluctuations and the
absence of zero momentum
correlations. Section VII describes the results of the
mean field and 1/N order spin correlations for the one
dimensional VBS
and GPFG states. {\em The most surprising result is that
for the VBS states of integer spin, the mean field, ${\cal O}(1/N)$
correction, and the exact result for N=2 are simply proportional!}
Section VIII summarizes what we have learned from our approach
in the context of large-N approximations in general.
It also lists some conjectures and open questions, which emerge from
this study. Appendices A,B, and C, fill in some technical details
which have been used to derive certain equations in the text.

\section{Schwinger Bosons Valence Bonds States}
\setcounter{equation}{0}

Schwinger bosons describe spin operators in a rotationally
invariant formulation. The standard SU(2) spin operators are given
by two commuting bosons $a,b$ at each site as follows:
\begin{eqnarray}
S_i^z~&=&\frac{1}{2} (a_i^\dagger a_i - b_i^\dagger b_i)\nonumber\\
S_i^+ &=& a_i^\dagger b_i\nonumber\\
S_i^- &=& b_i^\dagger a_i
\label{2a.1}\end{eqnarray}
The spin size $s$ is determined  by projecting the states
with the Gutzwiller operator
${\cal P}_s$ \cite{gut} onto the
subspace which obeys the local constraints at all sites
\begin{equation}
a_i^\dagger a_i + b_i^\dagger b_i~=n_{ai}+n_{bi}~=2s~.
\label{2a.2}\end{equation}
A Schwinger boson mean field wave function is defined as
\begin{equation}
|{\hat u }\rangle~= \exp\left[ \frac{1}{2} \sum_{ij} u_{ij}
(a^\dagger_i b^\dagger_j - b^\dagger_i a^\dagger_j )\right]|0\rangle
\label{2a.3}
\end{equation}
where $u_{ij}=-u_{ji}$ are either determined by some mean
field hamiltonian, or are
taken as free variational parameters. It is easy to verify that due
to the invariance of the forms $a^\dagger_i b^\dagger_j
- b^\dagger_i a^\dagger_j $ under global spin rotations,
$|{\hat u }\rangle$ is a total singlet. If we restrict $u_{ij}$ to
be {\em bipartite}, i.e. to connect only between two distinct
sublattices, say $A$ and $B$, then we
can redefine the operators on sublattice $B$ by sending
\begin{equation}
a_j\to -b_j~~,~~b_j\to a_j~~,~~~j\in B \label{2aaa.4}
\label{2a.3.1}
\end{equation}
and for $i\in A,~~j\in B,~~u_{ji}\to -u_{ji}$ and $u_{ij}\to
u_{ij}~$, so that ${\hat u }$ transforms into a symmetric matrix.
Under (\ref{2a.3.1}) the mean field wave function transforms into
\begin{equation}
|{\hat u }\rangle~\to ~\exp\left[ \frac{1}{2} \sum_{ij} u_{ij}
(a^\dagger_i a^\dagger_j + b^\dagger_i b^\dagger_j )\right]|0\rangle
\label{2a.4}
\end{equation}

Eqs. (\ref{2a.1}-\ref{2a.4}) can be generalized to SU(N)
representations using N flavors of Schwinger bosons
$a_{im},~m=0,\ldots ,N$. In order to construct symmetric forms
in the mean field wavefunction we again restrict
ourselves to bipartite lattices, and define the SU(N)
generalization $S_{imm'}$ of spin operators (\ref{2a.1}) as
\begin{equation}
{S}_{imm'}~\equiv~\cases{ a^\dagger_{im} a_{im'}&~$i\in A$\cr
                           -a^\dagger_{im'} a_{im}&~$i\in B$}
\label{2.3}
\end{equation}
where we have generalized the SU(2) sublattice rotation
(\ref{2aaa.4}) to SU(N).

The local constraints generalize to
\begin{equation}
\sum_{m=1}^N  n_{im}~\equiv~n_i~=Ns
\label{2.4}
\end{equation}
$Ns$ is an integer, where $s$ is a generalized ``spin size''.

The SU(N) generalization of our mean field wave function
(\ref{2a.4}) is
\begin{equation}
|\hat{u}\rangle ~=  \exp
\left[\frac{1}{2} \sum_{ij}u_{ij}\sum_{m=1}^{N}a^{\dag}_{im}
a^{\dag}_{jm}\right]|0\rangle.
\label{2.1}
\end{equation}
We list some essential properties of these states in Appendix A.

It is easy to show that  for any $m,m'$, and bond
$(i_a,j_b)$ we use the definitions (\ref{2.3}) and find that
\begin{eqnarray}
\left[ \sum_i {S}_{i m  m'},\sum_{\mu=1}^{N}a^{\dag}_{i_a \mu}
a^{\dag}_{j_b \mu}\right]  ~ &=& 0\nonumber\\
\sum_i {S}_{i m  m'}  |{\hat u}\rangle ~ &=&0.
\label{2.11.1}
\end{eqnarray}
Relations (\ref{2.11.1}) show that
$|\hat{u}\rangle$ is globally SU(N) invariant, and is therefore a
{\em singlet of total spin}.

We shall restrict ourselves to translationally
invariant states, which in Fourier
representation are given by
\begin{equation}
|\hat{u}\rangle ~=  \exp
\left[\frac{1}{2} \sum_{{\bf k }\in BZ} u_{{\bf k }}\sum_{m=1}^{N}a^
{\dag}_{{\bf k } m}a^{\dag}_{-{\bf k } m}\right]|0\rangle,
\label{2.12}
\end{equation}
where $BZ$ is the first Brillouin  zone,
$u_{\bf k }=\sum_j e^{i{\bf k }{ j }} u_{0j}$,
$ a^{\dag}_{{\bf k } m}= {\cal N}^{-\frac{1}{2}}\sum_j e^{i{\bf k }
{ j }}a^{\dag}_{jm}$, and ${\cal N}$ is the number of lattice sites.
We also define $S_{{\bf k } mm'}=\sum_je^{i{\bf k } j}S_{jmm'}$.

\subsection{The Gutzwiller Projection}

The mean field state (\ref{2.1}, \ref{2.12})
includes different spin sizes at each site. In order to construct a
bona-fide state of spins $s$, we must project out all other spin
sizes using the Gutzwiller projector
\begin{equation}
|{\hat u}\rangle_s~ ={\cal P}_{s} |{\hat u}\rangle
\label{2.7}
\end{equation}
which enforces the constraints (\ref{2.4}).
By expanding the exponential in (\ref{2.1})
and applying the Gutzwiller projection, we obtain
\begin{eqnarray}
|{\hat u}\rangle_s ~&=&{\cal P}_{s}~{1\over \nu_b !} ~ \left(
\frac{1}{2} \sum_{ij}u_{ij}\sum_{m=1}^{N}a^{\dag}_{im}
a^{\dag}_{jm}\right)^{\nu_b}|0\rangle\nonumber\\
 &=& {1\over \nu_b!} \sum_{\alpha}  \prod_{(ij)\in C_\alpha}
u_{ij}\left( \sum_{m=1}^{N}a^{\dag}_{im}
a^{\dag}_{jm}\right) |0\rangle\nonumber\\
\label{2.9}
\end{eqnarray}
where $\nu_b=\frac{1}{2}\,{\cal N} Ns$ is the total number of bonds
in the projected state and
$C_\alpha$ labels the different configurations of $\nu_b$ bonds on
the lattice, where exactly $Ns$ bonds eminate from every site.
In Fig. 1 we depict several configurations for various  $\{u_{ij}\}$
and values of $Ns$. $|{\hat u }\rangle_s$ which is a sum over such
configurations, is called a ``Valence Bonds State''.

Since all ${S}_{imm'}$ commute with the constraint,
$ |{\hat u}\rangle_s$ is also rotationally invariant and a total
singlet. If all the bond parameters are non negative, $u_{ij}\ge 0$,
the wavefunction (\ref{2.9}) satisfies the Marshall sign criterion
\cite{ms}. We recall that the ground state of any  bipartite
Heisenberg antiferromagnet must be a total singlet and obey
Marshall's theorem.

The correlation function in the Gutzwiller projected state is
defined as
\begin{equation}
S^{mm'}(i,j) = \langle S_{imm'} S_{jm'm} \rangle
\label{2.99r}
\end{equation}
and its Fourier transform is
\begin{eqnarray}
S^{mm'}({\bf k })~&=& \sum_j S^{mm'} ( 0, j ) \exp ( i {\bf k }
{\bf j} )
\nonumber \\
&=& {1 \over {\cal N}} \langle S_{-{\bf k } mm'} S_{{\bf k } m'm}
\rangle,
\label{2.99f}
\end{eqnarray}
where for any operator ${\cal O}$ we denote
\begin{equation}
\langle {\cal O} \rangle~\equiv \langle {\hat u }| {\cal P}_s
{\cal O} {\cal P}_s | {\hat u }
\rangle / \langle {\hat u }| {\cal P}_s | {\hat u } \rangle.
\end{equation}

\subsubsection{Resonating Valence Bonds ($s=\frac{1}{2}$)}

Several special cases have received particular attention in the
literature. For $N=2$, $s=1/2$ and nearest neighbors
$u_{\langle ij\rangle}$, $|{\hat u}\rangle_s$
is a superposition of all {\em dimer configurations}.
In one dimension there are only two such (Majumdar-Ghosh)
configurations \cite{mg}, which have an  exponentially small
overlap for large lattices. The spin correlation in one dimer state
vanishes beyond the nearest neighbor range.
In two and higher dimensions, $s=\frac{1}{2}$ states are sums over
many valence bonds configurations, which were denoted as
``Resonating Valence Bonds'' (RVB) by Anderson \cite{pwa}. One
configuration in the square lattice RVB state is depicted in
Fig. 1.a . The RVB state was proposed by Anderson \cite{pwa} and
others as trial ground states for frustrated quantum
antiferromagnets and high $T_c$ superconductors.
The number of dimer configurations grows exponentially with
${\cal N}$, and the overlap between different
configurations is finite.
The computation of the spin correlations in the dimer and longer
range RVB state was carried out numerically by Liang, Doucot and
Anderson \cite{lda} using bipartite bonds of various range.
They found that the RVB states with $u_{ij} \sim 1/r_{ij}^\alpha$
have long range order for  $\alpha\le 5$.

\subsubsection{Valence Bonds Solids, Integer $s$}

Affleck, Kennedy, Lieb and Tasaki \cite{aklt} (AKLT) found a class
of extended SU(2) Heisenberg models for which the exact ground
states are Valence Bond Solid (VBS) states, given for SU(N) by
\begin{eqnarray}
|\Psi^{vbs}\rangle_s~&=&\prod_{\langle i j\rangle} \left(\sum_{m=1}
^N a^\dagger_{im} a^\dagger_{jm}\right)^M |0\rangle,\nonumber\\
M~&=&Ns/z,
\label{2.13}
\end{eqnarray}
where $\langle i j\rangle$ denotes nearest neighbor bonds and
$z$ is the lattice coordination number. The condition that M
must be an integer restricts the size of the spin and
lattice for which such states can be defined (see Figs. 1.b, 1.c).
For example,
the SU(2) model in one dimension allows $s=1,2,3\ldots$.
On the square lattice, only even spins $s=2,4,6\ldots$ are
allowed. The correlation function $S^{+-}(i,j)\equiv \langle
\Psi^{vbs}|S^+_iS^-_j|\Psi^{vbs}\rangle/\langle \Psi^{vbs}|
\Psi^{vbs}\rangle$
for one dimension has been calculated for all $s$ by Arovas,
Auerbach and Haldane \cite{aah} (AAH) to be
\begin{eqnarray}
S^{+-}(i,j)~
&=& (-1)^{j-i} ~{2(s+1)^2 \over 3}~ \exp\left[  -\kappa_s |j-i|
\right]~ - \delta_{ij} {2(s+1)\over 3},\nonumber\\
\kappa_s~&=& \log(1+2/s),\nonumber\\
S^{+-}(k)~&=& {2 \over 3}(s+1)~{1-\cos(k)\over 1+\cos(k) +
{2\over s(s+2)}}.
\label{2.14}
\end{eqnarray}
The real space correlations decay as a pure exponential, with
$1/\kappa_s $ as the  correlation length. $|\Psi^{vbs}\rangle$ is a
``spin liquid'' ground state of the kind that was predicted
by Haldane \cite{hal} using the large-$s$, Non Linear Sigma model
analysis of the Heisenberg antiferromagnet. AAH also found
a Haldane gap \cite{hal} in its single mode excitation
spectrum.
The correlations
of VBS states on higher dimensional lattices are those of
a classical logarithmic Heisenberg model,  at temperature
$T=z/s$. This implies that for large enough $s$, the
VB states in three dimensions have long range N\'eel order.
The calculation of Eq. (\ref{2.14}) was performed in the
SU(2) coherent states basis. The generalization of this calculation
to $N>2$ has not yet been achieved. In the following, we shall apply
the large-N expansion to this problem.

It may be verified that the one dimensional Schwinger
boson state $|\hat{u}^{vbs}\rangle_s$ with
\begin{equation}
u^{vbs}_{ij}~=\delta_{\langle ij \rangle}~,~~~~~~
u^{vbs}_{k}~= 2 \cos(k)~,
\label{2.15}
\end{equation}
is dominated by the VBS state in the limit of infinite lattice
size ${\cal N}$:
\begin{eqnarray}
(M!)^{\frac{1}{2}{\cal N} z}~|{\hat u}^{vbs} \rangle_s~&=&
|\Psi^{vbs}\rangle_s~ +|\Psi'\rangle,
\label{2.16}
\end{eqnarray}
where $\langle\Psi'|\Psi'\rangle\sim c^{-{\cal N}}$
for some $c >1$. The exponentially small corrections are of non
uniform valence bonds configurations, where some bonds have
higher powers of $a^\dagger a^\dagger$ than others. Consequently,
we expect that in the thermodynamic limit
${\cal N}\rightarrow\infty$
the spin correlation function in the state $|\hat{u}^{vbs}
\rangle_s$ is given also by  (\ref{2.14}).

\section{Gutzwiller Projected Fermi Gas}
\setcounter{equation}{0}

In this section we introduce another important family of variational
states using fermions rather than Schwinger
bosons, to represent the SU(N) spin operators
\begin{equation}
{S}_{imm'}~\equiv~a^\dagger_{im} a_{im'}~,~~~~~~
a_{im}a^\dagger_{jm'}+a^\dagger_{jm'}a_{im}=\delta_{ij}
\delta_{mm'}.
\label{2b.1}
\end{equation}
The local constraint on the fermion occupation is
\begin{equation}
\sum_{m=1}^N  n_{im}~\equiv~n_i~=Ns,
\label{2b.2}
\end{equation}
where $Ns$ is an integer, which by the Pauli principle must be
less than or equal to $N$. Using the fermion operators, one can
construct a global SU(N) singlet by the following state
\begin{equation}
|{\hat u }\rangle_s~= {\cal P}_s \prod_{|{\bf k }| \le k_F}\left(
u_{\bf k } \prod_{m=1}^N a^\dagger_{{\bf k } m}\right)|0\rangle,
\label{2b.3}
\end{equation}
where $k_F$ is the Fermi momentum which is chosen to include $Ns$
states per site in the Fermi volume.  ${\cal P}_s$ Gutzwiller
projects onto the subspace which satisfies the
constraint (\ref{2b.2}). $u_{\bf k }$ are variational parameters.

It is possible to write (\ref{2b.3}) in an exponential form as
follows:
\begin{equation}
|{\hat u } \rangle_s~=
{\cal P}_{s}~  \exp \left[\frac{1}{2} \sum_{\bf k } u_{\bf k }
\theta\left(k_F-|{\bf k }|\right)~\sum_m a^\dagger_{{\bf k } m} a^
\dagger_{-{\bf k } m}\right]|0\rangle,
\label{2b.4}
\end{equation}
where $u_{\bf k }= -u_{-{\bf k }}$. In real space (\ref{2b.4}) are
analogous to the Schwinger boson states, defined in
(\ref{2.7}), (\ref{2.1}), where
$u_{ij}={\cal N}^{-1}\sum_{|{\bf k }|\le k_F}u_{{\bf k }}
e^{i{\bf k }(i-j)}$, $~~u_{ji}=-u_{ij}$. A well known case is the
Gutzwiller Projected Fermi Gas for $s=\frac{1}{2}$ (i.e. a
half-filled Brillouin zone),
\begin{equation}
|\Psi^{gpfg} \rangle~=
{\cal P}_{\frac{1}{2}}~  \exp \left[\frac{1}{2} \sum_{\bf k }
\mbox{sign}({\bf k })
\theta\left(k_F-|{\bf k }|\right)~\sum_m a^\dagger_{{\bf k } m} a^
\dagger_{-{\bf k } m}\right]|0\rangle.
\label{2b.5}
\end{equation}
In real space $|\Psi^{gpfg}\rangle$ contains long range bonds
$u_{ij},~ |i-j| >> 1$. Since the bonds are not bipartite, it does
not satisfy the Marshall sign criterion. This state is deduced from
the mean field theory of Baskaran, Zou and Anderson \cite{bza} for
the Heisenberg antiferromagnet. In one dimension,
$|\Psi^{gpfg}\rangle$ for SU(2) was found to be the exact ground
state of the Haldane-Shastry hamiltonian \cite{hs}, whose
interactions fall off as the second inverse power of distance.
This state has correlations, similar to that of the ground state of
the nearest neighbor Heisenberg model. Haldane
has also shown that the Haldane-Shastry hamiltonian and the
nearest neighbor Heisenberg model share similar gapless excitation
spectra \cite{hs}.

Gebhard and Vollhardt \cite{gv} have calculated the correlation
function of one dimensional $|\Psi^{gpfg}\rangle$ for N=2
\cite{note}:
\begin{equation}
S^{+-}(k)~ = -\frac{1}{2} \log \left( 1- {|k|\over \pi}\right)~.
\label{2.20}
\end{equation}
which in real space
decay asymptotically as a power law
\begin{equation}
S^{+-}(i,j)~={{\rm Si}[\pi (j-i)]\over 2\pi }~ {(-1)^
{(j-i)}\over j-i},
\label{mmm2}
\end{equation}
where ${\rm Si}(x)$ is the sine integral function.

\section{Correlations and the Generating Functional}
\setcounter{equation}{0}

The spin correlations of
$|{\hat u}\rangle_s$ can be derived from a generating functional.
The generating functionals for the
Valence Bonds states (\ref{2.7}) and the Gutzwiller Projected Fermi
Gas states (\ref{2b.4}) are formally very similar, and given by
\begin{equation}
Z[{ j }]~= \langle  ~\exp\left[ \sum_{imm'} \eta_ij_{imm'}
a^{\dag}_{im}a_{im'} \right] \rangle,
\label{3.1}
\end{equation}
where $a_{im}$ are either bosons or fermions, $j_{imm'}$ are the
source currents, $\eta_i=1$ for fermions and for bosons
\begin{equation}
\eta_i=\cases{1&$i\in A$\cr
-1&$i\in B$}~,
\end{equation}
which takes care of the sublattice rotation  of the
SU(N) spins (\ref{2.3}). The functional derivatives of
$Z$ determine the spin correlation functions.
It is sufficient to use
symmetric source matrices $j_{imm'}=j_{im'm}$. Hence,
$j_{imm'}$ and $j_{im'm}$ {\em are not to be considered as
independent}, but should be varied simultaneously when
differentiating $Z[j]$.

The following relations can be directly verified from (\ref{3.1})
\begin{equation}
\eta_i {\delta \log Z\over \delta { j }_{imm'}}\Bigg|_{{ j }=0}~
=\delta_{mm'}\langle n_{im} \rangle~=\delta_{mm'} s,
\label{3.2.1}
\end{equation}
which is a direct consequence of the SU(N) symmetry in
(\ref{2.1}) and the constraint (\ref{2.4}). The two-point spin
correlation function Eq. (\ref{2.99r}) is  given by
\begin{equation}
S^{mm'}(i,j)~= {1 \over 2 - \delta_{mm'}}\,Z^{-1} {\delta^2
Z\over \delta { j }_{imm'}\delta { j }_{jmm'}} \Bigg|_{{ j }=0}.
\label{3.2.2}
\end{equation}
Additional terms generated by the differentiation in
(\ref{3.2.2}) must vanish, since
\begin{equation}
\langle S_{i m \ne m'} S_{j m \ne m'} \rangle~ =0.
\label{3.2.3}
\end{equation}
Eq. (\ref{3.2.3}) follows from the rotational invariance of the wave
function. It is easy to verify, that
$S_{i m \ne m'} |{\hat u }\rangle_s$ and $S_{j m' \ne m} |{\hat u }
\rangle_s$
are eigenstates of the operator $\sum_i( S_{imm}- S_{im'm'})$ with
eigenvalues $+2$ and $-2$ respectively. Therefore these two states
are orthogonal and (\ref{3.2.3}) follows.

For $m\ne m'$, the correlation function $S^{m\ne m'}(i,j)$
is an SU(N) generalization of the usual SU(2) spin correlation
function $S^{+-}(i,j)=\langle S^+_iS^-_j\rangle$. In rotationally
invariant states this function is related to the correlations of
the other SU(2) spin components by
\begin{equation}
S^{m\ne m'}(i,j)~= 2\langle S^z_i S^z_j\rangle~={2\over 3}
\langle {\bf S }_i \cdot {\bf S }_j\rangle
\end{equation}

The evaluations of $Z$ and $S^{mm'}(i,j)$ of Eqs.
(\ref{3.1}, \ref{3.2.2}) are complicated because of the
Gutzwiller projector. If it were absent,  we could easily calculate
$Z$ as a matrix element of an exponential bilinear operator as done
in Appendix A. In order to proceed,
we must choose a convenient representation for the
projector. The projector can be represented as a limit of a
strongly interacting density matrix,
\begin{equation}
{\cal P}_{s}  ~ =
\lim_{\epsilon\to 0} \exp\left[ -{1\over 2N\epsilon^2}\sum_i
\left(n_i-Ns\right)^2\right]~.
\label{3.1a}
\end{equation}
Keeping $\epsilon$ finite will help to control infrared divergences
in subsequent diagrammatic calculations. The matrix elements of
(\ref{3.1a})
are hard to evaluate in its present form.
Using an auxiliary constraint variable $\lambda_i$ at every
site, we transform (\ref{3.1}) to an integral
\begin{eqnarray}
{\cal P}_{s} ~ &=& \lim_{\epsilon\to 0}  \int_{-\infty}^{\infty}
{\cal D} \lambda  ~\exp\left[ \sum_i \left(
-{N\epsilon^2 \over 2}\lambda_i^2 + i\lambda_i (n_i-Ns)\right)
\right],\nonumber\\
{\cal D} \lambda~&\equiv& \prod_i\left( \epsilon~\sqrt{N \over 2\pi}
d\lambda_i\right).
\label{3.2a}
\end{eqnarray}
Now we can write the generating functional as
\begin{eqnarray}
Z[{ j }]~&=&\lim_{\epsilon\to 0}  \int {\cal D} \lambda  ~\langle
{\hat u} |~
\exp\left[{\vec a}^\dagger ( i {\hat \lambda} + { \hat j} )
{\vec a} \right] ~|{\hat u} \rangle\nonumber\\
&&~~~~~~~~~~~~~~~~~\times ~e^{- iNs\sum_i   \lambda_i-{N\epsilon^2
\over 2}\sum_i \lambda_i^2  },
\label{3.3}
\end{eqnarray}
where we denote the matrices
\begin{equation}
\hat{\lambda}~=\lambda_i \delta_{ii'}\delta_{mm'}~,~~~~~
\hat{j}~= \eta_i j_{imm'} \delta_{ii'}~.
\end{equation}
In (\ref{3.3}) we have used the commutation of the spins with the
density operator,
\begin{equation}
\left[a^{\dag}_{im} a_{im'}\, ,\, n_{i'}\right]~=0
\label{3.4}
\end{equation}
to combine the exponentials of the source terms and the
projector. Now we use Appendix A, to evaluate (\ref{3.3}) as
\begin{eqnarray}
Z[{ j }]~&=&\lim_{\epsilon\to 0}  \int {\cal D} \lambda ~\exp\left(
N {\cal S}[\lambda , { j }] \right),\nonumber\\
 {\cal S}[\lambda, j]~&=&
 -{\zeta\over 2N} {\rm Tr}_{im}\log \left(1 - \zeta {\hat u^
\dagger} e^{i{\hat \lambda}+{\hat { j }} }{\hat u}  e^{i{\hat
\lambda}+{\hat { j }}} \right)\nonumber\\
&&~~~~~~~~~~~~~~~- i s \sum_i \lambda_i~-{\epsilon^2  \over 2}
\sum_i \lambda_i^2,
\label{3.5}
\end{eqnarray}
where $\zeta=+1~(-1)$ for bosons (fermions) and ${\hat u}^{\dag}$
is the hermitian conjugate of the matrix ${\hat u}$.

The correlation function is given by Eq. (\ref{3.2.2}):
\begin{eqnarray}
S^{mm'}(1,2)~&=& {1 \over 2 - \delta_{mm'}} \lim_{\epsilon\to 0}
Z^{-1} \int {\cal D} \lambda ~ \Biggl( N
{\delta^2 {\cal S}[\lambda, j] \over \delta { j }_{1mm'}\delta
{ j }_{2mm'}}\nonumber\\
&&~~~~~~~~~~~~~~~~~~~~+N^2\,
{\delta  {\cal S}[\lambda, j] \over \delta { j }_{1mm'}}\, {\delta
{\cal S}[\lambda, j] \over \delta { j }_{2mm'}}
\Biggr) \Bigg|_{j=0} ~\exp\left( N {\cal S}[\lambda]\right),
\nonumber\\
&&
\label{3.6}
\end{eqnarray}
where ${\cal S}[\lambda]~= {\cal S}[\lambda, j]\Big|_{j=0}$.

\section{The Diagrammatic Expansion of $Z(j)$}
\setcounter{equation}{0}

The multidimensional integration over ${\cal D}\lambda$
is equivalent to the difficult combinatorical problem of evaluating
the correlations in the valence bonds state, e.g. (\ref{2.9}). This
is clearly seen by expanding the action in powers of
$e^{i\lambda_i}$ and integrating with the weights
$e^{-iNs\lambda_i}$. The integrals reduce to
products of $\delta$-functions, which select the terms with
$Ns$ powers of $u_{ij}$ for any given site $i$. Summation over these
terms is  very cumbersome in general.

In (\ref{3.5},\ref{3.6}) the parameter $N$ was scaled out of the
action ${\cal S}$.  We shall evaluate the $\lambda$ integrals by
a saddle point expansion which is controlled
by the largeness of $N$.
The functional ${\cal S}[\lambda]$ is expanded as a Taylor series
about its minimum $\bar \lambda$; the coefficients of expansion
are independent of $N$.
Since the wave function is translationally
invariant we shall search in the space of {\em uniform} saddle
points ${\bar \lambda_i}={\bar \lambda}$. ${\bar\lambda}$
is found by requiring that the linear variations vanish.  We define
$e^{i{\bar \lambda}}~={\bar u}$. The saddle point or ``mean
field'' equation is
\begin{equation}
{ 1 \over i }{ \delta {\cal S}[\lambda] \over \delta \lambda_j }
\Bigg|_{\bar{\lambda}}~=\left[{\bar u}^2\hat{u}^{\dag} \hat{u}
( 1 - \zeta
{\bar u}^2 \hat{u}^{\dag} \hat{u} )^{-1} \right]_{jj} - s~=0,
\label{3.8a}
\end{equation}
which determines ${\bar u}[{\hat u},S]$. We see that
(\ref{3.8a}) implies that ${\bar u}$ is real, i.e. the
integration paths of the variables  $\lambda_i$ have to be
analytically continued to cross the imaginary axis
at $\bar\lambda~=-i\ln({\bar u})$. The term $[~]_{jj}$
in (\ref{3.8a}) is the average number of particles of every
flavor in the (unconstrained) state $|\bar{u}\hat{u}\rangle$ (see
Appendix A). Thus, the mean field equation yields that the
constraint is satisfied on the average. If we define the unprojected
state using ${\bar u}{\hat u}$ instead of $\hat{u}$,
the mean field equation (\ref{3.8a}) would be satisfied with
$\bar{\lambda}=0$.  Hence we shall
use that convention for ${\hat u}$ and set $\bar\lambda=0$.

We now expand ${\cal S}$, in powers of $\lambda$ to obtain
\begin{equation}
{\cal S}[\lambda]~= {\cal S}^{(0)}~ - \frac{1}{2} \sum_{ij}
{\cal S}^{(2)}_{ij} \lambda_i \lambda_j~ + {\cal S}^{\rm int},
\label{3.9aa}
\end{equation}
where ${\cal S}^{\rm int}$ includes only third and higher
order terms:
\begin{equation}
{\cal S}^{\rm int}~= \sum_{n=3}^\infty ~{i^n \over n!}
{\cal S}^{(n)}_{i_1,\ldots, i_n}~\lambda_{i_1}\cdots  \lambda_{i_n}.
\label{3.9a}
\end{equation}
${\cal S}^{(n)}$ here depend parametrically on ${\bar u}$.
Diagrammatically they are depicted as thick circles with $n$
$\lambda$ vertices denoted by wiggly lines (see Fig. 2 ).
Later we shall obtain explicit expressions for ${\cal S}^{(n)}$ in
form of {\em loops} constructed of Greens functions.

We also expand the pre-exponential functions in (\ref{3.6})
\begin{eqnarray}
{\delta {\cal S}[\lambda,j] \over \delta j_{1 mm'}}\,\Bigg|_{j=0}
&=& {\delta_{mm'}\eta_1\over N} \left( \sum_{n=1}^\infty {i^n \over
n!}\, {\cal S}^{(n+1)}_{1,i_1,\ldots, i_n}  \lambda_{i_1}\cdots
\lambda_{i_n} + s \right) \nonumber\\
&=& { \delta_{mm'}\eta_1 \over N } \sum_{n=0}^{\infty} { i^n \over
n! }\, {\cal S}^{(n+1)}_{1,i_1,\ldots,i_n} \lambda_{i_1} \cdots
\lambda_{i_n}, \nonumber \\
{1 \over 2 - \delta_{mm'}} {\delta^2 {\cal S}[\lambda,j] \over\delta
j_{1mm'} \delta j_{2mm'} }\,\Bigg|_{j=0}
&=& {\eta_1\eta_2\over N} \sum_{n=0}^\infty { i^n \over n! }\,
{\cal S}^{(n+2)}_{1,2,i_1,\ldots, i_n}~  \lambda_{i_1}\cdots
\lambda_{i_n},
\label{3.11a}
\end{eqnarray}
where sums over repeated indices are assumed and we denoted
${\cal S}_1^{(1)}= s$. Here we use (\ref{3.5}) to relate the
derivatives with respect to $\lambda$ to those with respect to $j$.
Diagrammatically, we denote the current vertices (at points $1,2$)
by dashed lines as shown in Fig. 2.

The correlations functions can be evaluated by inserting the
expansions (\ref{3.9aa}) and (\ref{3.11a}) into (\ref{3.6})
\begin{equation}
S^{mm'}(1,2)~= \eta_1 \eta_2 \left[  S^{I}(1,2) ~+ \delta_{mm'}
S^{II}(1,2) \right] ~,
\label{3.11.0}
\end{equation}
where
\begin{eqnarray}
S^{I}(1,2) &=&
\lim_{\epsilon\to 0} Z^{-1} \int {\cal D} \lambda ~\left(
\sum_{n=0}^\infty { i^n \over n! }\, {\cal S}^{(n+2)}_{1,2,i_1,
\ldots , i_n}\lambda_{i_1} \cdots \lambda_{i_n}\right)\nonumber\\
&&~~~~~\times
\sum_{L=0}^\infty {N^L \over L!} \left( \sum_{n=3}^\infty ~
{ i^n \over n! }\, {\cal S}^{(n)}_{i_1,\ldots ,i_n}~  \lambda_{i_1}
\cdots  \lambda_{i_n} \right)^L~\exp\left( -{N \over 2}
{\cal S}^{(2)}_{kl}\lambda_{k} \lambda_{l} \right),\nonumber\\
S^{II}(1,2) &=&
\lim_{\epsilon\to 0} Z^{-1} \int {\cal D} \lambda \left(
\sum_{n=0}^\infty {i^n \over n!}\,{\cal S}^{(n+1)}_{1,i_1,\ldots ,
i_n}\lambda_{i_1}\cdots \lambda_{i_n} \right) \left(
\sum_{n=0}^\infty { i^n \over n! }\, {\cal S}^{(n+1)}_{2,i_1,
\ldots ,i_n} \lambda_{i_1}\cdots \lambda_{i_n} \right)\nonumber\\
&&~~~~~\times \sum_{L=0}^\infty {N^L \over L!} \left( \sum_{n=3}^
\infty ~{i^n \over n!}\, {\cal S}^{(n)}_{i_1,\ldots, i_n}~  \lambda_
{i_1}\cdots \lambda_{i_n}\right)^L~\exp\left( -{N\over 2}
{\cal S}^{(2)}_{kl} \lambda_{k} \lambda_{l} \right).\nonumber\\
\label{3.11}
\end{eqnarray}
We disregard the contribution of the constant $N{\cal S}^{(0)}$
in the action.

The integrals in (\ref{3.11}) are sums of
multidimensional gaussian integrals.
The gaussian integrations  contract all $\lambda$ fields in
pairs  $\overbrace{\lambda_k\lambda_l}$ bringing down a propagator
for each pair given by
\begin{equation}
D_{kl}~= \overbrace{\lambda_k\lambda_l}~= - {1\over N} \left(
S^{(2)}\right)^{-1}_{kl},
\label{3.12}
\end{equation}
where the minus sign is due to the factor of $i$, which accompanies
every $\lambda$ field. The propagator (\ref{3.12}) is depicted as a
wavyline connecting two $\lambda$ vertices in Fig. 2.
One has to sum over all $\lambda$
contractions. The disconnected parts of the diagrams serve to
cancel the factor of $Z^{-1}$, leaving us with the diagrams which
are connected to one or the other current vertices (A linked
cluster theorem).
Thus calculating any particular diagram involves multiplying loops
${\cal S}^{(n)}$ and propagators $D$, and summing over internal
lattice points. There are {\em internal} loops, created by powers of
${\cal S}^{\rm int}$, and {\em external} loops, coming from an
expansion of the preexponential functions (\ref{3.11a}). Internal
loops must have at least three $\lambda$ vertices; external loops
have current vertices and might also have arbitrary (including
zero) number of $\lambda$ vertices. The order of any particular
diagram is given by
\begin{equation}
\left({1\over N} \right)^{P-L}
\label{3.13}
\end{equation}
where $L$ is the number of internal loops (or the power $L$
of the sums in (\ref{3.11})), and $P$ is the number of propagators
(half the number of $\lambda$ fields in (\ref{3.11})). After
grouping all the diagrams at each order in 1/N we obtain the series
\begin{equation}
S^{mm'}(1,2)~=\sum_{p=0}^\infty N^{-p} S^{mm'\,(p)}(1,2).
\label{3.13a}
\end{equation}

Similar rules govern the calculation of higher correlation
functions. One has to sum over all
possible ways of distributing the current vertices on the
external loops. Within each loop the $m,m'$-indices of the external
currents must be equal to the indices of other external currents, to
allow non zero values of the trace.

\section{Identities To All Orders}
\setcounter{equation}{0}

{\em The diagrammatic expansion of the $1/N$ series has special
structure which allow us to obtain exact identities to
all orders in $1/N$}. A key feature is that the propagator $D$ of
the constraint field $\lambda$ is non other than the inverse of the
square part of the action ${\cal S}^{(2)}$. As a result we shall
show that the local constraints are exactly enforced
to each order in the expansion, i.e. there are no contributions
to charge fluctuations when all terms of the same order are
considered. In addition we shall prove a sum rule for the on site
spin fluctuations for arbitrary $N$ and the absence of
zero momentum off-diagonal spin correlations.

\subsection{Absence of Charge Fluctuations}

Here we shall demonstrate that the constraint is imposed
at each order in $1/N$. In other words, due to the Gutzwiller
projection the density fluctuations vanish identically after
all diagrams of a given order are summed, yielding
\begin{equation}
\langle n_1 {\cal A} \rangle
{}~=Ns~\langle {\cal A} \rangle
\label{3.19}
\end{equation}
for an arbitrary operator ${\cal A}$. It is instructive to
see how (\ref{3.19}) is derived by the diagrammatic
expansion. A current of $n_1$ belongs to some external loop
${\cal S}^{(n+1)},\,\, n\ge 0$. Let us first consider all the
contributions
with $n\ge 1$. We define a {\em ``tail''} of a diagram as the
combination of a propagator attached in series to a loop $S^{(2)}$
which has the operator $n_1$ on its other vertex.
All diagrams can be separated into two classes: ones with a tail,
and ones without a tail. It is easy to identify for each diagram
without a tail say $R(n_1,{\cal A})$, a counterterm
${\bar R}(n_1,{\cal A})$ by attaching a tail to the $n_1$
vertex (See Fig. 3). By (\ref{3.12}) the two are of the
same order $p$ (they have the same number of loops minus
propagators) and they cancel precisely
\begin{eqnarray}
\bar{R}(n_1,{\cal A})~&=&N\sum_{kl}{\cal S}^{(2)}_{1,k}D_{{kl}}
R(n_l,{\cal A}) \nonumber\\
{}~&=&-R(n_1,{\cal A}).
\label{3.21}
\end{eqnarray}
Thus, at any order $p$, the
counterterms cancel the connected charge fluctuation diagrams
one by one. The only terms in the expansion of $\langle
n_1{\cal A}\rangle$ which survive are the disconnected contributions
with $n_1$ on the loop ${\cal S}^{(1)}$. An important property of
the diagram rules is the absence of the counterterm to the
${\cal S}^{(1)}$ loop. Such counterterm would involve an internal
loop ${\cal S}^{(1)}$, which is not allowed by the rules of our
expansion. Thus (\ref{3.19}) follows from ${\cal S}^{(1)}=s$. Q.E.D.

\subsection{Sum Rule for the On-Site Spin Fluctuations}

First we shall show that for SU(N) invariant states there is
a relation between two types of
correlation functions: diagonal $S^{mm}$ and off-diagonal
$S^{m\ne m'}$. Both functions do not depend on the particular
values of $m,m'$ since the wave functions are SU(N) symmetric.
Due to the $\delta_{mm'}$ coefficient in (\ref{3.11.0}),
$S^{m\ne m'}$ is equal to $S^{I}$. Let $S_c^{mm}$ be the
connected part of the diagonal correlation function.
Any diagram which contributes to the off-diagonal correlations
also appears in  $S_c^{mm}$, i.e.
\begin{equation}
{\rm if}~~~
R^{\alpha}~\in~S^{m\ne m'}~,~~~{\rm then}~~~R^{\alpha}~\in~S_c^{mm}.
\end{equation}
However, in $S_c^{mm}$ its contribution is partially cancelled
by the counterterms, which are given by tails insertion. It may be
seen, that the sum of any diagram $R^{\alpha}\in S^{m\ne m'}$ and
its counterterms, obtained by all possible ways of insertion of
tails, is equal to $(1-1/N)R^{\alpha}$.

In addition to diagrams of the type $R^{\alpha}$ and their
counterterms, yet another contribution to $S_c^{mm}$ is given by
the diagrams $R^\beta$, which have the two current vertices on
different loops ${\cal S}^{(n_1+1)}$ and ${\cal S}^{(n_2+1)},\,\,
n_{1,2}\ge 2$, and by their counterterms $\bar{R}^\beta$. But the
diagrams $R^\beta$ must exactly cancel with their counterterms! This
is verified by adding a tail to one of the loops and seeing that
the counterterm is of the same order in 1/N  due to the
additional $m$-summation for the loop which became an internal
loop. This proves that all the diagrams which do not cancel are
of the type $R^{\alpha}$ and the following important identity holds:
\begin{equation}
S_c^{mm}(i,j)~=\left( 1-{1\over N}\right)~S^{m\ne m'}(i,j).
\label{3.22'}
\end{equation}

Now we can calculate  the onsite spin fluctuations.
For SU(2) we are familiar with
the ``spin square'' operator ${\bf S }^2$, which when projected to
the $s$ sector yields a diagonal matrix of elements $s(s+1)$.
For larger $N$, its natural generalization is
\begin{eqnarray}
{\bf S }_i^2~&\equiv&{\cal P}_{Ns} \sum_{mm'} {S}_{imm'} {S}_{im'm}
\nonumber\\
{}~&=&{\cal P}_{Ns}~\sum_{mm'} \left( n_{im'} (1~+\zeta n_{im})-\zeta
\delta_{mm'}n_{im'}\right)\nonumber\\
{}~&=& \left( N^2 s(1+\zeta s) -\zeta N\right)~ {\cal P}_{Ns},
\label{3.22}
\end{eqnarray}
where in the fermion case we made use of equality $n_{im}^2=
n_{im}\,\,(\zeta=-1)$. On the other hand, using (\ref{3.22'})
we can write the onsite fluctuations as
\begin{eqnarray}
\langle {\bf S }_i^2\rangle~&=&\sum_{m\ne m'}S^{m\ne m'}(i,i)~+
\sum_{m}\left( S_c^{mm}(i,i)+\langle n_{im}\rangle^2\right)
\nonumber\\
{}~&=&\left( N(N-1)+N(1-{1\over N}) \right)S^{m\ne m'}(i,i)~+Ns^2.
\label{3.26}
\end{eqnarray}
Using (\ref{3.22}) we find the desired identity
\begin{equation}
S^{m\ne m'}(i,i)~= {N\over N+\zeta}~s(1+\zeta s).
\label{3.27}
\end{equation}
For SU(2) this formula reduces to the known values:
\begin{equation}
\langle S^+_i S^-_i\rangle~=\cases{ {2\over 3} s(s+1)&Schwinger
bosons\cr
\frac{1}{2} & fermions ($s$=1/2)}
\label{3.28}
\end{equation}
In translationally invariant cases (\ref{3.27}) gives a sum rule,
which is useful for checking the diagrammatic calculations at each
order of 1/N seperately. In momentum space,
the diagrams of order $(1/N)^p$ must obey
\begin{equation}
{\cal N}^{-1} \sum_{\bf k } S^{m\ne m' (p)}({\bf k })~= \left(
{1\over N}\right)^p ~(-\zeta)^p s(1+\zeta s),~~~~~~~p=0,1,2,\ldots
\label{3.30}
\end{equation}

\subsection{Absence of zero momentum correlations}

The last identity is a consequence of the singlet nature
of the wavefunction which implies that $S_{{\bf k }=0\, m\ne m'}
|{\hat u }\rangle_s = 0$. By using Eq. (\ref{2.99f}) we obtain from
that
\begin{equation}
S^{m\ne m'}({\bf k }=0)~=0~,
\label{3.32}
\end{equation}
which holds of course at each and every order in the $1/N$ series.

\section{Calculations of leading orders}
\setcounter{equation}{0}

In this section we calculate  the spin correlation functions
$S^{m\ne m'}$ using the
$1/N$ expansion.  We start with an explicit evaluation of
the loops ${\cal S}^{(n)}$ and propagator $D$. In the cases of
interest $\hat{u}$ is hermitian. It is useful to define
the following {\em Greens functions}
\begin{equation}
\hat{u}_{\pm}~=\zeta^{1/2} \bar{u} \hat{u} \left( \pm 1 -
\zeta^{1/2} \bar{u} \hat{u} \right)^{-1}.
\label{mmm7.3}
\end{equation}
We also introduce the matrix $\Lambda=\delta_{ij}(e^{i\lambda_j}-1)$
and express the action (\ref{3.5}) as
\begin{eqnarray}
{\cal S}[\lambda]~&=& {\cal S}^{(0)} - {\zeta \over 2} \sum_
{\gamma=\pm}{\rm Tr} \log ( 1 - \hat{u}_{\gamma} \Lambda ) - i s
\sum_i \lambda_i - {\epsilon^2 \over 2} \sum_i \lambda_i^2,
\nonumber\\
{\cal S}^{(0)}~&=& -\,{\zeta \over 2} \sum_{\gamma=\pm} {\rm Tr}
\log ( 1 - \gamma \zeta^{1/2} {\bar u} \hat{u} ).
\label{mmm7.4}
\end{eqnarray}
${\cal S}^{(0)}$ is a constant which we shall disregard.

Expanding the logarithm in (\ref{mmm7.4}) and using
(\ref{3.8a}) to cancel the linear term we obtain
\begin{equation}
{\cal S}[\lambda]~= {\zeta \over 2} {\sum_{\gamma,\, n=1}^{\infty}}'
{1 \over n} {\rm Tr}\,( \hat{u}_{\gamma} \Lambda )^n -
{\epsilon^2 \over 2} \sum_j \lambda_j^2,
\label{mmm7.4a}
\end{equation}
where  ${\sum}'$ denotes that terms linear in
$\lambda$ are excluded. By equating terms of the same order in
$\lambda$ in Eqs. (\ref{3.9aa}) and (\ref{mmm7.4a}) we can relate
the loops ${\cal S}^{(n)},\,n\ge 2$ to traces over Greens functions
$u_\gamma$. Diagrammatically, we denote the Greens functions by
thin sold lines. A closed loop of Greens functions denotes a
trace over lattice and $\gamma$ indices. For a $n$ $\lambda$
vertices ${\cal S}^{(n)}$, there are contributions from diagrams
with $1\le m\le n$ Greens functions, since the function $\Lambda$
yields all powers of $\lambda$ fields at the same point. Due to
Eq. (\ref{mmm10.1}) below, the loop ${\cal S}^{(1)}$ may be denoted
diagrammatically in the same manner -- as a closed loop of one
Greens function with one vertex.

The translational invariance of ${\hat u}$ makes it
easier to work in the momentum representation. The linear action,
or the mean field equation (\ref{3.8a}), is explicitly given by
\begin{equation}
{\zeta \over 2}\, ( \hat{u}_+ + \hat{u}_- )_{jj} ~~=
{1 \over {\cal N}}\, \sum_{{\bf k }} {\bar u}^2 {|u_{{\bf k }}|^2
\over 1 - \zeta {\bar u}^2|u_{{\bf k }}|^2} ~= s.
\label{mmm10.1}
\end{equation}
The quadratic action in (\ref{3.9aa}) is given by
\begin{eqnarray}
{\cal S}^{(2)}_{ij}~&=&{\zeta \over 2}\sum_{\gamma}
( {u_{\gamma}}_{ij}
{u_{\gamma}}_{ji} + \delta_{ij} {u_{\gamma}}_{ii} ) + \delta_{ij}
\epsilon^2    \nonumber\\
&=&{\zeta \over 2} \sum_{\gamma} {u_{\gamma}}_{ij} {u_{\gamma}}_
{ji} + \delta_{ij} s + \delta_{ij} \epsilon^2, \nonumber\\
{\cal S}^{(2)}({\bf q })~&=&{\zeta \over 2 {\cal N}}\,
\sum_{{\bf k }} \sum_{\gamma=\pm} \, {u_{\gamma}}_{{\bf k }}\,
{u_{\gamma}}_{{\bf k } + {\bf q }} + s + \epsilon^2 \nonumber\\
&=&{1 \over 2 {\cal N}}\, \sum_{{\bf k }} \sum_{\gamma=\pm 1}\,
{ {\bar u}^2 u_{{\bf k }}\,u_{{\bf k } + {\bf q }} \over ( \gamma -
\zeta^{1/2} {\bar u}u_{{\bf k }} )\,( \gamma - \zeta^{1/2}{\bar u}
u_{{\bf k }+{\bf q }} )} + s + \epsilon^2,
\label{mmm10.2}
\end{eqnarray}
and the propagator is
\begin{equation}
D({\bf q })=-\,\left( N {\cal S}^{(2)}({\bf q }) \right)^{-1}.
\label{mmm10.3}
\end{equation}
An important property of this expansion is that the lowest order
(mean field) correlation function $S^{m\neq m'\,(0)}({\bf q })$ is
simply related to the quadratic part of the action:
\begin{eqnarray}
S^{m\neq m'(0)}(i,j)~&=&\eta_i \eta_j {\cal S}^{(2)}(i,j),
\nonumber \\
S^{m\neq m' \, (0)}({\bf q })~&=&\cases{{\cal S}^{(2)}({\bf q } +
\vec{\pi}), & Schwinger bosons \cr
{\cal S}^{(2)}({\bf q }), & fermions}
\label{mmm11.1}
\end{eqnarray}
where $\vec{\pi}=(\pi,\,\pi,\, \ldots)$ for a cubic lattice.
The diagrams for $S^{m\neq m'\,(0)}({\bf q })$ are shown in Fig. 4.
At this point we note that $D({\bf q })$ is singular for $\epsilon
\to 0$ since $S^{m\neq m' \, (0)}({\bf q }=0)$ vanishes by
(\ref{3.32}). This causes diagrams which involve one or more
propagators to diverge as $1/\epsilon$. A check
on the correctness of the calculation is that these ``infrared''
divergences must exactly cancel between different diagrams to
yield a finite result for $\lim_{\epsilon\to 0} S^{(p)}(1,2)$ for
each order $p$ separately. We shall come back to this point in our
summary.

The 1/N corrections for $S^{m\neq m'}({\bf k })$ are given by the
diagrams of Fig. 5. Solid lines represent factors of
$u_\gamma(k)$. Each vertex conserves momentum, and
indices $\gamma=+,-$ of the solid lines.
We must sum over internal momenta and  $\gamma$.
Diagrams with external currents 1,2 at the same point
denote an overall factor of $\delta_{12}$. Due to the cancellation
mechanism described in section VI.1, the third and fourth diagrams
in the bottom row of Fig. 5, cancel against the fifth diagram. Thus
by the same mechanism, there is complete cancellation between the
last four diagrams in the second row. We shall describe the
calculations of the remaining diagrams
for the Scwhinger boson and fermion cases separately,
and defer technical details to Appendices B and C.

\subsection{Valence Bonds Solid Correlations}

The mean field equation Eq. (\ref{mmm10.1}) for the Valence Bonds
Solid state (\ref{2.15}) is
\begin{equation}
\int_{-\pi}^{\pi} {dk \over 2\pi}~{4 \bar{u}^2 \cos^2(k) \over
1-4 \bar{u}^2 \cos^2(k)}~=s,
\label{4.1}
\end{equation}
whose solution is
\begin{equation}
\bar{u}(s)~={\sqrt{s(s+2)}\over 2(s+1)}~.
\label{4.2}
\end{equation}
By (\ref{mmm11.1}) and (\ref{mmm10.2}) we obtain
\begin{equation}
S^{m\neq m' (0)}(k) =- 2 \bar{u}^2 \sum_{\gamma=\pm1} \int_{-\pi}
^{\pi} {dq \over 2 \pi} { \cos(q) \cos(k + q) \over ( \gamma -
2 \bar{u} \cos(q) ) ( \gamma + 2 \bar{u} \cos(k + q) ) } + s.
\label{mmm15.1}
\end{equation}
The integral is performed by introducing a new variable $z=e^{iq}$
which transforms the integration over $q$ into an integration
along a unit circle $|z|=1$ in the complex $z$-plane. Using
(\ref{4.2}) yields
\begin{equation}
S^{m\ne m' (0)} (k)= (s+1)~{1-\cos(k)\over 1+\cos(k) +
{2 \over s(s+2)}}.
\label{4.3}
\end{equation}
This result is very  surprising, since it is just
proportional to the exact result for $N=2$ as found  by Ref.
\cite{aah} and given in
Eq. (\ref{2.14}). The factor $2/3$ between (\ref{2.14}) and
the mean field result (\ref{4.3}) is consistent with the factor
$N/(N+1)$ between the mean field on site fluctuations
and the exact sum rule, Eq. (\ref{3.27}).
This suggests that perhaps the exact
$N$ dependence of $S^{m\ne m'}(k)$ is given by this
simple multiplicative factor. Fortunately, we are able to
calculate the  ${\cal O}(1/N)$ corrections and check this
proposition at least to the next leading order
in $1/N$. This calculation
is described by the diagrams of Fig. 5, which involve exchanges of
one or two propagators $D$. The sum of all diagrams was evaluated
analytically using the symbolic manipulation program Mathematica
\cite{math}. The result, derived in Appendix B, is
\begin{equation}
S^{m\ne m' (1)} (k)~= -{1\over N} S^{m\ne m' (0)} (k).
\label{4.4}
\end{equation}
This result confirms the above hypothesis, but is far from obvious!
In fact, the separate 1/N diagrams have infra-red divergences of
order $\epsilon^{-1}$ due to the diverging
propagators at momentum $\pi$. In addition, the separate diagrams
have different correlation lengths than the mean field function, but
these effects somehow cancel by summing {\em all} the terms of order
1/N, leaving us with  Eq. (\ref{4.4}). It is highly tempting to
conjecture that the same relation holds to all orders, i.e. that
\begin{equation}
S^{m\ne m'\,(p)} (k)~= \left(-{1\over N} \right)^p S^{m\ne m'(0)}
(k)~~~(?)
\label{4.5}
\end{equation}
which will sum up to the simple relation
\begin{equation}
S^{m\ne m'} (k)~= {N\over N+1} ~ S^{m\ne m' (0)} (k)~~~(?)
\label{4.6}
\end{equation}
For N=2, we have already seen that this conjecture is correct.
But,  as we shall discuss in section VIII, {\it the
underlying reason for this relation is still a mystery.}

\subsection{Fermions GPFG Correlations}

The mean field equation (\ref{mmm10.1}) for the one dimensional
GPFG state $|\Psi^{gpfg}\rangle$ (\ref{2b.5}) is given by the
integral
\begin{equation}
{\bar{u}^2\over 1+\bar{u}^2}\int_{-\pi}^{\pi} {dk \over 2\pi}
\theta(k_F-|k|) ~-s~= 0,
\label{5.1}
\end{equation}
with $k_F=\pi/2$ and $s=1/2$. Eq. (\ref{5.1}) yields
\begin{equation}
\bar{u}(s)~={1\over \sqrt{{k_F\over \pi s} -1}}
\label{5.2}
\end{equation}
which implies that for $k_F\to\pi/2$ and  $s=1/2$,
${\bar u}\to \infty$. To enable us to calculate the correlations
for the GPFG, we keep ${\bar u}$ finite by
holding a Fermi level $k_F$ a little above $\pi/2$.
We shall take the limit $k_F\to\pi/2$ only at the end of our
calculations. This divergence simplifies the calculations
considerably, because Eq. (\ref{mmm7.3}) yields a simple limit
\begin{equation}
\lim_{k_F\to{\pi\over 2}} u_{\pm}(q)=-\theta\left({\pi\over 2}
- |q| \right).
\label{mmm17.2}
\end{equation}
The lowest order correlation function is evaluated using
(\ref{mmm11.1}), (\ref{mmm10.2}) and (\ref{mmm17.2})
\begin{eqnarray}
\lim_{k_F\to{\pi\over 2}} S^{m\ne m' (0)} (k)~&=&\frac{1}{2} -
\int_{-\pi}^{\pi} {dq \over 2\pi} \theta\left({\pi\over 2}-|q|
\right) \theta\left({\pi\over 2}-|k+q|\right)
\nonumber\\
{}~&=&{|k|\over 2\pi}
\label{5.3}
\end{eqnarray}

When comparing (\ref{5.3}) to Gebhard and Vollhardt's
result, Eq. (\ref{2.20}), we find that the two
expressions agree very well in the small $k$ limit, where they
vanish with the same linear coefficient,
but they deviate at larger $k$ as shown in Fig. 6. Eq. (\ref{2.20})
diverges logarithmically near $k=\pi$,  while the mean field result
(\ref{5.3}) has merely
a discontinuity in its derivative. This translates to a difference
in the asymptotic power law decay in real space, between $1/|i-j|$
of Eq. (\ref{mmm2}) and $1/|i-j|^2$ of (\ref{5.3}).
There is a factor of $N/(N-1)=2$ between their sum rules
as required by Eq. (\ref{3.27}).

In Appendix C, we calculate the  $1/N$ diagrams for the
GPFG state. We obtain the result:
\begin{equation}
S^{m\ne m' (1)} (k)~= {1\over N}\left[ {|k|\over \pi}+ \left( 1-{|k|
\over \pi} \right) \log \left( 1-{|k|\over \pi} \right) \right].
\label{5.4}
\end{equation}
In Fig. 6 we compare the functions $S^{m\ne m'(0)} (k)$,
 $S^{m\ne m'(0)} (k)+ S^{m\ne m'(1)} (k)$ and the exact result
Eq. (\ref{2.20}) for N=2.
We see that the $1/N$ correction improves the mean field
approximation considerably near the zone boundary, where its
derivative diverges logarithmically. In real space we obtain for
separations $r=|j-i|$,
\begin{equation}
S^{m\ne m' (1)} (r)~= {(-1)^r\over N \pi^2 r^2 }\left[\gamma+
\log(\pi r) - {\rm Ci}(\pi r) \right]
\label{5.4a}
\end{equation}
where $\gamma=0.577\ldots$ is the Euler constant, and $ {\rm Ci}(x)
{}~=-\int_x^\infty dt \cos(t)/t$ vanishes for large $x$. In
(\ref{5.4a}) we find that the 1/N correction enhances the long
distance correlations from $r^{-2}$ to $r^{-2} \log(r)$.

\section{Summary and Discussion}
\setcounter{equation}{0}

In this paper we have introduced a  large-N expansion for the
correlation functions of Gutzwiller projected states. We have
discovered several properties and sum rules which hold to all
orders in $1/N$.
By explicitly calculating the mean field and 1/N corrections
for particular Schwinger boson and fermion states, we can
check the validity of this approach against exact results for N=2.
We shall conclude by discussing what we believe we have learned from
our results, and what still needs to be illuminated by further
investigations and insight.

\subsection{What We Understand}

\begin{enumerate}

\item The effects of the Gutzwiller projector can be expanded
systematically in terms
of $1/N$ diagrams. Each diagram with $L$ loops and $P$ propagators
is of order $(1/N)^{(P-L)}$. The loops and propagators are
determined by the mean field (saddle pont) equation.

\item The local charge fluctuations are suppressed at each order by
counterterms, which have the tail structure depicted in Fig. 3.

\item For any $N$, the onsite spin fluctuations are given by
\begin{equation}
\langle S_{im\ne m'} S_{im'\ne m} \rangle~= {N\over N+\zeta} s
(1+\zeta s)
\label{8.1}
\end{equation}
where $\zeta=1$ $(-1)$ for bosons (fermions).

\item Each diagram can diverge due to the divergence of the
propagators (see discussion after Eq. (\ref{mmm11.1}). However,
the sum of all diagrams of the 1/N order
is finite. We conclude that in general, large-N expansions are prone
to such intermediate divergences, due to the ``hardness'' of the
constraints (or lack of ``self-interaction'' for the $\lambda$
fields). The lesson to be learned is that {\em results which are
based on any subset of diagrams, or on partial resummation
schemes, are highly suspect.}

\end{enumerate}

\subsection{What We Do Not Understand}

\begin{enumerate}

\item  For the VBS states, the mean field, ${\cal O}(1/N)$
and the exact $N=2$ correlations are simply proportional.
We conjecture that the higher order terms behave in the same manner,
i.e.
\begin{equation}
S^{m\ne m'} (k)~= {N\over N+1} ~ S^{m\ne m' (0)} (k)~~~(?)
\label{4.6aaa}
\end{equation}
For Schwinger bosons, we know that this relation holds for the
on-site sum rule (\ref{8.1}),
but its validity for all $k$ is a surprise. We can recall however
that similar surprises have been found in other large-N
calculations, both with bosons and with fermions, where
mean field results differ from the exact result
by the  factor $N/(N+\zeta)$. For example:
The mean field  susceptibilities of the $s=\frac{1}{2}$ ferromagnet
in one dimension, and antiferromagnet in two dimensions \cite{aa}
(both for $N=2$) are off by a factor of $2/3=N(N+1)$. Also, the
Wilson ratios of the Kondo impurity model \cite{ki} and the
$s=\frac{1}{2}$ Heisenberg antiferromagnetic
chain \cite{aa} are $2=N/(N-1)$. It would therefore be very useful
to understand this relation in order to  correct the mean field
approximation for other problems. The apparent simplicity of
this correction factor may have its origin in some group theoretical
relation between the saddle point approximation and exact integrals
over Haar measures \cite{MS}.

\item The above discussion indicates that for these systems the 1/N
expansion is not just an asymptotic series but a convergent,
well behaved expansion. On the other hand we are faced with the
apparent failure of the boson large-N theory  for the Valence Bonds
Solid at $s=\frac{1}{2}$ in one dimension.
The $1/N$ series yields exponentially decaying correlations, while
the correct state (the nearest neighbor dimers state)
has vanishing correlations beyond nearest neighbor
separations. We therefore strongly
suspect that there is an essential singularity in the expansion of
the form
\begin{equation}
\frac{1}{2} \left[ 1 + \exp( i 2 \pi N s)\right]
\end{equation}
which cannot be obtained at any order in the expansion. Such a
factor distinguishes between integer and half odd integer spins for
$N=2$. This is similar to
the famous topological Berry's phase, or ``$\theta$-term'',
of the continuom theory  of half odd integer Heisenberg
antiferromagnets in one dimension. This term must be added to the
Schwinger boson mean field Lagrangian to obtain the correct ground
state degeneracies \cite{RS}.

\item We note that the Fermion large-N approximation is quite
successful for the $s=\frac{1}{2}$ GPFG state in one dimension. The
$1/N$ corrections enhance the long distance correlations from
$r^{-2}$ to $r^{-2}\log(r)$. It would be interesting to find out how
the full $1/N$ series modifies the power law to $r^{-1}$ for $N=2$.

We recall that the fermion mean field theory for the spin half
Heisenberg chain \cite{bza} is successful in reproducing the
Fermi-liquid features of the exact solution \cite{aa}.
Here we have found another empirical evidence that the fermionic
approach is better than the bosonic approach for $s=\frac{1}{2}$
antiferromagnets in one dimension. In two dimensions, the relative
advantage of the fermionic versus bosonic large-N approach is not
clear.

\end{enumerate}

\subsection*{Acknowledgements}

This work has been supported by the National Science Foundation,
NSF-DMR-9213884,
and grant number 90-0041/1 from the US-Israel Binational Science
Foundation.

\setcounter{section}{0}
\renewcommand{\thesection}{Appendix \Alph{section}:}
\renewcommand{\theequation}{\Alph{section}\arabic{equation}}

\section{The Properties of Mean Field States}
\setcounter{equation}{0}

The major goal of this appendix is the derivation of Eq.
(\ref{3.5}). We will define here the mean field states $|\hat{u}
\rangle$ by
\begin{equation}
| \hat{u} \rangle = \exp \left[ \frac{1}{2} \sum_{ij} u_{ij}
a_i^{\dag} a_j^{\dag} \right] | 0 \rangle,
\label{a1.1}
\end{equation}
where $a_i$ are either bosons or fermions, satisfying the usual
commutate relations
\begin{equation}
a_i a^{\dag}_j - \zeta a^{\dag}_j a_i = \delta_{ij},~~~~~
a_i a_j - \zeta a_j a_i = a^{\dag}_i a^{\dag}_j - \zeta a^{\dag}_j
a^{\dag}_i = 0,
\end{equation}
$\zeta=+1~(-1)$ for bosons (fermions) and matrix $\hat u$ satisfies
the symmetry condition $\hat{u}^T=\zeta \hat{u}\;\;(u^T_{ij}=
u_{ji})$. We have dropped the $m$ indices for simplicity. The states
$|\hat u \rangle$ have the following important properties:

\begin{enumerate}

\item The overlap of mean field states is given by
\begin{equation}
\langle \hat{u} | \hat{v} \rangle = \left[ \det ( 1 - \zeta
\hat{u}^{\dag} \hat{v} ) \right]^{- {\zeta \over 2}}.
\label{a1.2}
\end{equation}

\item Let $A$ is an arbitrary operator, the product of any number of
creation and annihilation operators. Then the extended Wick's
theorem holds: the normalized matrix element of $A$, $\langle
\hat{u}|A|\hat{v}\rangle/\langle \hat{u}|\hat{v}\rangle$, is equal
to the sum of all possible completly contracted products of creation
and anihilation operators (with the usual sign for fermions), in
which each contraction involves two operators. A contraction of
$d_1,\, d_2$ is the normalized expectation value
$\langle \hat{u}|d_1d_2|\hat{v}\rangle/\langle\hat{u}|\hat{v}
\rangle$.

\item The contractions are given by
\begin{equation}
\begin{array}{ccccl}
{ \langle \hat{u} | a_i a_j | \hat{v} \rangle \over \langle
\hat{u} | \hat{v} \rangle } &=& \zeta \left[ \hat{v} ( 1 - \zeta
\hat{u}^{\dag} \hat{v} )^{ - 1 } \right]_{ ij } &=& \left(
{ \langle \hat{v} | a_j^{\dag} a_i^{\dag} | \hat{u} \rangle \over
\langle \hat{v} | \hat{u} \rangle } \right)^{\ast} , \\

{ \langle \hat{u} | a_i^{\dag} a_j | \hat{v} \rangle \over \langle
\hat{u} | \hat{v} \rangle } &=& \left[ \hat{u}^{\dag} \hat{v}( 1-
\zeta \hat{u}^{\dag} \hat{v} )^{ - 1 } \right]_{ij} &=& \zeta\,
{ \langle \hat{u} | a_j a_i^{\dag} | \hat{v} \rangle \over \langle
\hat{u} | \hat{v} \rangle } - \zeta \delta_{ij}\, .
\end{array}
\label{a1.3}
\end{equation}
\end{enumerate}
The proof of these properties may be found in \cite{bb} (see also
\cite{rs}).

Following \cite{bb} we will write the whole set of creation and
annihilation operators as a $2{\cal N}$-dimensional vector
(${\cal N}$ is the lattice size):
\begin{equation}
\vec{\gamma} \equiv \{ \vec{a}, \vec{a}^{\dag} \} \equiv \{ a_1,
\ldots , a_{{\cal N}}, a_1^{\dag}, \ldots , a_{{\cal N}}^{\dag} \}
\label{a2.1}
\end{equation}
with commutation relations
\begin{equation}
\gamma_i \gamma_j - \zeta \gamma_j \gamma_i = \rho_{ij},
\label{a2.2}
\end{equation}
where $\rho$ is the $2{\cal N}\times 2{\cal N}$ matrix
\begin{equation}
\rho = \left( \begin{array}{cc} 0 & 1 \\ - \zeta & 0 \end{array}
\right).
\label{a2.3}
\end{equation}
We have $\rho^2=-\zeta$, so $\rho^{-1}=-\zeta\rho$. Following Balian
and Brezin \cite{bb} we define
\begin{equation}
{\cal J} = \exp \left[ \frac{1}{2} \vec{\gamma} R \vec{\gamma}
\right] \equiv \exp \left[ \frac{1}{2} \sum_{i,j=1}^{2 {\cal N}}
\gamma_i R_{ij} \gamma_j \right],
\label{a2.4}
\end{equation}
where the $2{\cal N}\times2{\cal N}$ matrix $R$ satisfies the
symmetry condition
$R^T=\zeta R$ (in this appendix we will call such matrices
symmetric) and $2{\cal N}\times2{\cal N}$ matrices
\begin{equation}
T \equiv \left( \begin{array}{cc} T_{11} & T_{12} \\ T_{21} &
T_{22} \end{array} \right) = e^{ \rho R}.
\label{a2.5}
\end{equation}
It is shown in \cite{bb}, that the matrices (\ref{a2.5}) faithfully
represent the second quantized operators (\ref{a2.4}), i.e.
\begin{equation}
{\cal J}(R_1)
{\cal J}(R_2)={\cal J}(R)~~\Rightarrow~~
e^{\rho R_1}e^{\rho R_2}=
e^{\rho R}~.
\label{a2.5a}
\end{equation}
We will now prove the inverse statement,
\begin{equation}
e^{\rho R_1}e^{\rho R_2}=
e^{\rho R}~~\Rightarrow~~
{\cal J}(R_1){\cal J}(R_2)={\cal J}(R)~,
\label{a2.5b}
\end{equation}
that is to say, the representation of operators $\cal J$ by
matrices $T$ is isomorphic. To prove the (\ref{a2.5b}), we use the
Baker -- Campbell -- Hausdorff formula, which gives \cite{j}
\begin{equation}
e^{c_1} e^{c_2} = \exp \left( \sum_{n=1}^{\infty} \sum_{l_1,\ldots
,l_n=1,2} \alpha_{l_1,\ldots,l_n} [ c_{l_n} [ c_{l_{n-1}} \cdots
[ c_{l_2}, c_{l_1} ] \cdots ]] \right),
\label{a3.1}
\end{equation}
where we denoted $c_{1,2}=\frac{1}{2} \vec{\gamma}R_{1,2}
\vec{\gamma}$
and $\alpha_{l_1,\ldots,l_n}$ are some constants, which we do not
need to know explicitly. On the other hand, since $e^{\rho R_1}
e^{\rho R_2}=e^{\rho R}$, the same formula gives
\begin{equation}
R = \rho^{-1} \sum_{n=1}^{\infty} \sum_{l_1,\ldots,l_n=1,2}
\alpha_{l_1,\ldots,l_n} [ \rho R_{l_n} [ \rho R_{l_{n-1}} \cdots
[ \rho R_{l_2}, \rho R_{l_1} ] \cdots ]]\,.
\label{a3.2}
\end{equation}
Correspondingly, (\ref{a2.5b}) will be proven, if one can show that
\begin{equation}
[ c_{l_n} [ c_{l_{n-1}} \cdots [ c_{l_2}, c_{l_1} ] \cdots ]] =
\frac{1}{2} \vec{\gamma} \rho^{-1} [ \rho R_{l_n} [ \rho R_{l_{n-1}}
\cdots [ \rho R_{l_2}, \rho R_{l_1} ] \cdots ]] \vec{\gamma}\,.
\label{a3.3}
\end{equation}
Let us denote
\begin{equation}
A_n = \rho^{-1} [ \rho R_{l_n} [ \rho R_{l_{n-1}} \cdots [ \rho
R_{l_2}, \rho R_{l_1} ] \cdots ]]\,,
\label{a3.4}
\end{equation}
so that the r.h.s. of (\ref{a3.3}) is equal to $\frac{1}{2}
\vec{\gamma}A_n\vec{\gamma}$. We will first prove by induction, that
$A_n$ is symmetric, $A_n^T=\zeta A_n$. For $n=1$ it is correct;
then, for $A_{n+1}$ we have
\begin{equation}
A_{n+1} = \rho^{-1} [ \rho R_{l_{n+1}}, \rho A_n ] = R_{l_{n+1}}
\rho A_n - A_n \rho R_{l_{n+1}}\,,
\label{a4.1}
\end{equation}
and since $R_{l_{n+1}}$ is symmetric and $\rho$ is antisymmetric,
$\rho^T=-\zeta\rho$, the symmetry of $A_{n+1}$ follows from
the symmetry of $A_n$.

Now we will prove by induction the relation (\ref{a3.3}). For
$n=1$ it is trivially correct; then, if it is correct for $n$, we
have for $n+1$:
\begin{equation}
[ c_{l_{n+1}} [ c_{l_n} \cdots [ c_{l_2}, c_{l_1} ] \cdots ]] =
[ \frac{1}{2} \vec{\gamma} R_{l_{n+1}} \vec{\gamma}, \frac{1}{2}
\vec{\gamma}A_n \vec{\gamma} ].
\label{a4.2}
\end{equation}
Using symmetry of $R_{l_{n+1}}$ and $A_n$ and the commutation
relation for $\gamma$ (\ref{a2.2}) it is straightforward to show,
that r.h.s. of (\ref{a4.2}) is equal to
\begin{equation}
\frac{1}{2} \vec{\gamma} \rho^{-1} [ \rho R_{l_{n+1}}, \rho A_n ]
\vec{\gamma} = \frac{1}{2} \vec{\gamma} \rho^{-1} [ \rho R_{l_{n+1}}
[ \rho R_{l_n} \cdots [ \rho R_{l_2}, \rho R_{l_1} ] \cdots ]]
\vec{\gamma},
\label{a4.3}
\end{equation}
which completes the proof of (\ref{a2.5b}). Note, that we have also
shown, that the product of two operators of the type (\ref{a2.4})
is another operator of the same type, represented by the matrix $R$
of (\ref{a3.2}).

Now we shall use this ``multiplication rule''  to calculate
(\ref{3.3}). The expectation value is of the form
\begin{equation}
M = \langle \hat{u} | \exp \left[ \vec{a}^{\dag} \hat{\eta}
\vec{a} \right] | \hat{u} \rangle = \langle 0 | {\cal J}_1 \exp
\left[ \vec{a}^{\dag} \hat{\eta} \vec{a} \right] {\cal J}_2 | 0
\rangle,
\label{a5.1}
\end{equation}
where ${\cal J}_1=\exp\left[\frac{1}{2}\vec{a}\hat{u}^{\dag}\vec{a}
\right]$ and ${\cal J}_2=\exp\left[\frac{1}{2}\vec{a}^{\dag}\hat{u}
\vec{a}^{\dag}\right]$ are operators of the type (\ref{a2.4}) and
$\hat{\eta}$ is ${\cal N}\times{\cal N}$ matrix. We can transform
$\exp\left[\vec{a}^{\dag}\hat{\eta}\vec{a}\right]$ also into the
form (\ref{a2.4}) by writing
\begin{equation}
\exp \left[ \vec{a}^{\dag} \hat{\eta} \vec{a} \right] = \exp
\left[ -{ \zeta \over 2}\mbox{Tr}\, \hat{\eta} \right] \exp \left[
\frac{1}{2} \vec{\gamma} \left( \begin{array}{cc}
0 & \zeta \hat{\eta}^T \\ \hat{\eta} & 0 \end{array} \right)
\vec{\gamma} \right] \equiv \exp \left[ - { \zeta \over 2 }
\mbox{Tr}\, \hat{\eta} \right] {\cal J}_3,
\label{a5.2}
\end{equation}
where ${\cal J}_3$ is of the type (\ref{a2.4}), so we have for $M$
\begin{eqnarray}
M &=& \exp \left[ - { \zeta \over 2 } \mbox{Tr}\, \hat{\eta}
\right] \langle 0 | {\cal J}_1 {\cal J}_3 {\cal J}_2 | 0 \rangle
\nonumber \\
&=& \left[ \det e^{\hat{\eta}} \right]^{ - { \zeta \over 2 } }
\langle 0 | {\cal J}_1 {\cal J}_3 {\cal J}_2 | 0 \rangle.
\label{a5.3}
\end{eqnarray}
The $T$ matrices, corresponding to ${\cal J}_1$, ${\cal J}_2$ and
${\cal J}_3$ are
\begin{equation}
\begin{array}{cclcl}
T_1 &=& \exp \left[ \left( \begin{array}{cc} 0 & 1 \\ - \zeta & 0
\end{array} \right) \left( \begin{array}{cc} \hat{u}^{\dag} & 0 \\
0 & 0 \end{array} \right) \right] &=& \left( \begin{array}{cc} 1 &
0 \\ - \zeta \hat{u}^{\dag} & 1 \end{array} \right), \\
 & & & & \\
T_2 &=& \exp \left[ \left( \begin{array}{cc} 0 & 1 \\ - \zeta & 0
\end{array} \right) \left( \begin{array}{cc} 0 & 0 \\ 0 & \hat{u}
\end{array} \right) \right] &=& \left( \begin{array}{cc} 1 &
\hat{u} \\ 0 & 1 \end{array} \right), \\
 & & & & \\
T_3 &=& \exp \left[ \left( \begin{array}{cc} 0 & 1 \\ -\zeta & 0
\end{array} \right) \left( \begin{array}{cc} 0 & \zeta\hat{\eta}^T
\\ \hat{\eta} & 0 \end{array} \right) \right] &=& \left(
\begin{array}{cc} e^{\hat{\eta}} & 0 \\ 0 & e^{-\hat{\eta}^T}
\end{array} \right),
\end{array}
\label{a5.4}
\end{equation}
Thus,  $T$, which corresponds to ${\cal J}={\cal J}_1{\cal
J}_3{\cal J}_2$ is
\begin{equation}
T = T_1 T_3 T_2 = \left( \begin{array}{cc} e^{\hat{\eta}} &\,\,
e^{\hat{\eta}}\,\hat{u} \\-\zeta\, \hat{u}^{\dag} e^{\hat{\eta}} &
\;\;\;\;e^{-\hat{\eta}^T} - \zeta\, \hat{u}^{\dag} e^{\hat{\eta}}\,
\hat{u} \end{array} \right).
\label{a6.1}
\end{equation}
It is shown in \cite{bb}, that for every operator $\cal J$
(\ref{a2.4}) and corresponding matrix $T$ the following relation
holds:
\begin{equation}
\langle 0 | {\cal J} | 0 \rangle = \left[ \det T_{22} \right]
^{-{\zeta\over2}}.
\label{a6.2}
\end{equation}
Using Eqs. (\ref{a6.2}), (\ref{a6.1}) and (\ref{a5.3}) we will
obtain for the case of {\it symmetric} (in usual sense) $\hat
{\eta}$, $\hat{\eta}^T=\hat{\eta}$, the final relation
\begin{eqnarray}
\langle \hat{u} | \exp \left[ \vec{a}^{\dag} \hat{\eta} \vec{a}
\right] | \hat{u} \rangle &=& \left[ \det \left( 1 - \zeta\,
\hat{u}^{\dag} e^{\hat{\eta}}\, \hat{u} e^{\hat{\eta}} \right)
\right]^{-{\zeta\over2}} \nonumber \\
&=& \exp \left[ - { \zeta \over 2 } {\rm Tr} \ln \left( 1 - \zeta
\,\hat{u}^{\dag} e^{\hat{\eta}}\, \hat{u} e^{\hat{\eta}} \right)
\right],
\label{a6.3}
\end{eqnarray}
which yields the expression (\ref{3.5}) for the generating
functional (\ref{3.3}).

\section{Calculation of (\protect\ref{4.4})}
\setcounter{equation}{0}

Here we explicitly calculate the $1/N$ order diagrams
for the Schwinger boson case. The  diagrams
are depicted in Fig. 5.  We use  the integration
variable $z=e^{ik}$ instead of the momentum $k$.
For example, for the nearest neighbor bonds problem
$\hat{u}^{vbs}$ of (\ref{2.15}), we have
\begin{equation}
u^{vbs}(z)~=z+\frac{1}{z}\,.     \label{b1.1}
\end{equation}
The conservation of momentum at every vertex is equivalent to
the rule that the product of all $z$'s entering a vertex
is equal to unity. Each sum
over $k$ \cite{comm1} is replaced in the thermodynamic limit
by a contour integration over $z$ on the unit circle
\begin{eqnarray}
\lim_{{\cal N}\to \infty} {\cal N}^{-1}\sum_k F_k~=\frac{1}{2\pi}
\int_{-\pi}^{\pi} dk F_k~&\longrightarrow&\frac{1}
{2\pi i}\oint_{|z|=1}\frac{dz}{z}F(z)\nonumber\\
&=&\sum_i {\rm Res}\left[F(z_i)/z_i\right]
\end{eqnarray}
For the quadratic part of the action (\ref{mmm10.2}) the integral is
\begin{equation}
{\cal S}^{(2)}(z)=\frac{1}{2}\sum_{\gamma}\frac{1}{2\pi i}\oint\frac
{dz'}{z'}u_{\gamma}(z')u_{\gamma}(zz')+s+\epsilon^2,
\label{b1.2}
\end{equation}
where
\begin{equation}
u_\pm (z)=\bar{u} u(z) \left[ \pm 1 - \bar{u} u(z) \right]^{-1}
\label{b1.3}
\end{equation}
and for the valence bond case $u(z)$ is given by (\ref{b1.1}) and
$\bar{u}$ by (\ref{4.2}).

Since trigonometric integrands $F_k$
are replaced by rational functions  $F(z)$, it is easy to determine
their poles $z_i$ (including a pole at $z=0$), and their residues
at these poles. The sum over residues in (\ref{b1.2}) yields
\begin{equation}
{\cal S}^{(2)}(z)=(s+1)\kappa\frac{(z+1)^2}{(z-\kappa)(1-z\kappa)}+
\epsilon^2,
\label{b1.4}
\end{equation}
where
\begin{equation}
\kappa=\frac{s}{s+2}\,.
\label{b1.5}
\end{equation}
The propagator $D(z)$ is equal to $-\left(N{\cal S}^{(2)}(z)\right)^
{-1}$. The diagrams are generated by the rules of Section 3, and
shown in Fig. 5. As an example,  the integrations of the vertex
diagram (see Fig. 7) are
\begin{equation}
\frac{1}{2}\sum_{\gamma}\frac{1}{(2\pi i)^2}\oint
\frac{dz'}{z'}\oint\frac{dz''}{dz''} ~u_{\gamma}(z')~u_{\gamma}(zz')
{}~u_{\gamma}(zz'z'') ~u_{\gamma}(z'z'')~D(z'')~~.
\label{b1.6}
\end{equation}

Each diagram in Fig. 5. diverges as $1/\epsilon $. However,  the
divergences cancel in the sum, and the overall 1/N  result is
finite for $\epsilon\to 0$. The simplicity of
the residues method allowed us to use the symbolic
manipulation program ``Mathematica'' \cite{math} to perform the
integrations analytically on the computer. The program identifies
the poles and residues of the rational functions.
Intermediate expressions, especially for the diagram Fig.~7,
involved upto hundreds of terms.
Expanding these terms and finding common denominators became
too cumbersome for manual calculations, and therefore
automating this process was essential.
The result of this calculation is given by Eq.(\ref{4.4}).

\section{Calculation of (\protect\ref{5.4})}
\setcounter{equation}{0}

Here we derive the order 1/N correlations of the Gutzwiller
Projected Fermi Gas state, Eq.(\ref{5.4}).
The quadratic part of the action is given by
(\ref{5.3}), so by (\ref{mmm10.3}), we have a diverging propagator
at $k=0$. We control this divergence by letting $k_F>\pi/2$.
We denote
\begin{equation}
 { 2 k_F \over \pi } - 1~\equiv { \delta \over 2 \pi } > 0~.
\label{c1.1}
\end{equation}
The role of $\delta$  is similar to that of $\epsilon$ -- it
regulates the divergence of the propagator. Thus we obtain
\begin{equation}
{\cal S}^{(2)}(k) = { \delta + |k| \over 2 \pi }\,,
\label{c1.2}
\end{equation}
so that
\begin{equation}
D(k) ~= - { 1 \over N } { 2 \pi \over \delta + |k| }\,.
\label{c1.3}
\end{equation}
We first note that the
last two diagrams of the top row in Fig. 5, cancel since they
differ by one $u_{\gamma}$ line, which yields a factor of
$-1$. For the same reason, the sum of the first two diagrams
is equal to twice the contribution of the
second diagram. The remaining contributions
to $S^{m\neq m' (1)}(k)$ are depicted (including combinatorial
factors) in Fig. 8. By reflection symmery, we can restrict ourselves
to  $k>0$.

The first diagram of Fig. 8 is given by
\begin{equation}
R_1(k) = \int_{-\pi}^{\pi} D(q) P(k+q) { dq \over 2 \pi },
\label{c2.1}
\end{equation}
where $P(k)$ is the polarization bubble. By Fig. 4, for $k$ in the
first Brillouin zone
\begin{equation}
P(k) = {\cal S}^{(2)}(k) - s = { |k| \over 2 \pi } - \frac{1}{2} \,,
\label{c2.2}
\end{equation}
which yields for (\ref{c2.1})
\begin{equation}
R_1(k) = \frac{1}{N} \left[ \left( 1 - \frac{k}{\pi} \right) \log
\frac{\pi-k}
{\delta} - \frac{k}{\pi} \log \frac{k}{\pi} + \frac{2k}{\pi}-1
\right]\,.
\label{c2.3}
\end{equation}
The contribution of the second diagram of Fig. 8 is
\begin{equation}
2 R_2(k) = 2 \int_{-\pi}^{\pi} { dq \over 2 \pi } D(q) \int_{-\pi}
^{\pi} { dp \over 2 \pi }\, \theta\, ( \frac{\pi}{2} - |p| )\,
\theta\, ( \frac{\pi}{2} - |p+q| )\, \theta\, ( \frac{\pi}{2}
- |p+q+k| )\,.
\label{c2.5}
\end{equation}
We denote the integral over $p$ in (\ref{c2.5}) as
$(2\pi)^{-1}A(q)$, where
\begin{equation}
A(q) = \left\{ \begin{array}{lll}
 \pi+q & \mbox{ if } - \pi \leq q \leq -k \\
 \pi-k & \mbox{ if } - k \leq q \leq 0 \\
 ( \pi - q - k )\, \theta\, ( \pi - q - k ) & \mbox{ if } 0 \leq q
 \leq \pi \end{array} \right.
\label{c2.6}
\end{equation}
which yields
\begin{equation}
2 R_2(k)= \frac{1}{N} \left[ -\frac{1}{\pi} \int_
{-\pi}^{-k} \frac{\pi+q}{-q} dq - \frac{1}{\pi} \int_{-k}^0
\frac{\pi-k}{\delta-q} dq - \frac{1}{\pi} \int_0^{\pi-k}
\frac{\pi-q-k}{\delta+q} dq \right]\,.
\label{c2.7}
\end{equation}
The last diagram of Fig. 8 is
\begin{eqnarray}
R_3(k) &=& - \int_{-\pi}^{\pi} { dq \over 2 \pi } D(q) \int_{-\pi}
^{\pi} { dp \over 2 \pi }\, \theta\, ( \frac{\pi}{2} - |p| )\,
\theta\, ( \frac{\pi}{2} - |p+k| ) \nonumber \\
&& \qquad\qquad\qquad \times \theta\, ( \frac{\pi}{2} - |p+q| )\,
\theta \,( \frac{\pi}{2} - |p+k+q| )\,.
\label{c2.8}
\end{eqnarray}
The integral over $p$ is equal to $(2\pi)^{-1}(\pi-k-|q|)\,\theta\,
(\pi-k-|q|)$, so (\ref{c2.8}) cancels the last term of
(\ref{c2.7}), while first two terms of (\ref{c2.7}) yield
\begin{equation}
2 R_2(k) + R_3(k) =
\frac{1}{N} \left( \log \frac{\delta}{\pi} - \frac{k}{\pi} \log
\frac{\delta}{k} + 1 - \frac{k}{\pi} \right)\,.
\label{c2.9}
\end{equation}
By adding (\ref{c2.9}) and (\ref{c2.3}) we obtain Eq. (\ref{5.4}).

\vfill\eject
\parindent=0pt
\underbar{\bf Figure Captions:}

\bigskip

{\bf Fig. 1}: Graphical representation of valence bonds
configurations, which contribute to (a) Resonating Valence Bonds
States (RVB) on the square lattice.
(b) Valence Bonds Solid on the square lattice. (c) Valence Bonds
Solid on the chain.
\medskip

{\bf Fig. 2}: Diagrammatic representation of the 1/N expansion for
the \linebreak correlation function $S(1,2)$. Thick circles denote
loops, wiggly lines are \mbox{$\lambda$-propagators,} and dashed
lines are external currents.
\medskip

{\bf Fig. 3}: Diagrammatic representation of cancellation of charge
fluctuations by counterterms, Eq. (\ref{3.21}).
\medskip

{\bf Fig. 4}: The diagrams contributing to the loop
${\cal S}^{(2)}$. $u_\pm$ are
mean field Greens functions, defined in Eq. (\ref{mmm7.3}).
\medskip

{\bf Fig. 5}: The 1/N corrections to $S^{m\ne m'}({\bf k })$. See
discussion after Eq. (\ref{mmm11.1}).
\medskip

{\bf Fig. 6}: The N=2 spin correlations of the Guztwiller Projected
Fermi Gas in one dimension. The exact result (solid line, from Ref.
\cite{gv}) diverges at $k=\pi$.
The mean field (MF, short dashes) result has discontinuos
derivatives at $k=0,\pi$, while the sum up to order 1/N (long dashed
line) has a diverging derivative at $k=\pi$ (see Eq. (\ref{5.4})).
\medskip

{\bf Fig. 7}: Assignments of integration variables in
Eq. (\ref{b1.6}).
\medskip

{\bf Fig. 8}: Diagrams contributing to the correlations of the GPFG
(Appendix~C).


\begin{thebibliography}{99}
\bibitem{note} Note, that definition of spin operators in
\cite{gv} is different from our's (\ref{2a.1}) by a factor of 2, so
we divided their spin correlation function by 4 to put it in
agreement with our notations.
\bibitem{comm1} The substitution of the momenta summation by an
integration
is correct in the thermodynamic (${\cal N}\to \infty$) limit,
provided the integrand $F(k)$ is finite
for all $k$. In the case of long range magnetic order, special care
has to be taken of ``Bose condensates'' at $k=k^{bc}$ where
$F(k^{bc}) \propto {\cal N}$.
\end{thebibliography}
\end{document}